\documentclass[aps,prd,preprint,groupedaddress]{revtex4-1}

 \usepackage{graphicx} 

\begin{document}

\title{  Bottomonium dissociation in a finite density plasma    }



\author{Nelson R. F. Braga}\email{braga@if.ufrj.br}
\affiliation{Instituto de F\'{\i}sica,
Universidade Federal do Rio de Janeiro, Caixa Postal 68528, RJ
21941-972 -- Brazil}

\author{Luiz F.  Ferreira}
\email{luizfaulhaber@if.ufrj.br}
\affiliation{Instituto de F\'{\i}sica,
Universidade Federal do Rio de Janeiro, Caixa Postal 68528, RJ
21941-972 -- Brazil}


\begin{abstract}
 We present a holographic description of the thermal behavior  of $ b \bar b$ heavy vector mesons   inside a plasma at finite temperature and density.  The meson dissociation in the medium 
is represented by the decrease in the height of the spectral function peaks. 
In order to find a description for the evolution of the quasi-states with temperature and chemical potential it is crucial to use a model that is consistent with  the decay constant behavior. 
The reason is that the height of a spectral function  peak  is related to   the value of the zero temperature decay constant of the corresponding particle. 
AdS/QCD holographic models are in general not consistent with the observation that decay constants of heavy vector mesons decrease with radial excitation level.
However, it was recently shown that using a soft wall background and  calculating the correlation functions at a finite  position of anti-de Sitter space, associated with an ultraviolet energy scale,  it is possible to describe the observed behavior. 
Here we extend this proposal to the case of finite temperature $T $ and chemical potential $\mu $. A clear picture of the dissociation of bottomonium states as a function of $ \mu $ and $T$  emerges from the spectral function. The energy scales where the change in chemical potential leads to changes in the thermal properties of the mesons is consistent with QCD expectations. 
           
\end{abstract}

\keywords{Gauge-gravity correspondence, Phenomenological Models}

\maketitle

\section{ Introduction }   

Understanding the thermal properties of heavy vector mesons inside a plasma of quarks and gluons can be a helpful tool for investigating heavy ion collisions. 
The dissociation of such particles may indicate the formation of a thermal medium. 
This type of proposal, considering charmonium states,  appeared  a long time ago in \cite{Matsui:1986dk} (see also \cite{Satz:2005hx}). 

A holographic description of the dissociation of charmonium and bottomonium states in a plasma at finite temperature but  zero density appeared recently in ref. \cite{Braga:2016wkm}.  In this article the first radial excitations $1S, 2S$ and $ 3S$ appear as clear peaks of the spectral function. The height of the peaks  decrease as the temperature of the medium increases, as expected.  This reference used a finite temperature version of the  holographic AdS/QCD model proposed in ref. \cite{Braga:2015jca} for calculating decay constants and masses of vector mesons. For completeness we mention that a previous model that describes the thermal behavior of the first excited state of charmonium appeared before in 
ref. \cite{Grigoryan:2010pj}.

AdS/QCD models,  inspired in the AdS/CFT correspondence \cite{Maldacena:1997re,Gubser:1998bc,Witten:1998qj},  provide nice fits for hadronic mass spectra.  
The simplest one is the hard wall  AdS/QCD model,  proposed in refs. \cite{Polchinski:2001tt,BoschiFilho:2002ta,BoschiFilho:2002vd,deTeramond:2005su}
and consists in placing a hard cutoff in anti-de Sitter (AdS)  space.   
 Another AdS/QCD model is the soft wall that has the  property that the square of the mass  grow linearly with the radial excitation number \cite{Karch:2006pv}. In this case the background involves AdS  space and a scalar field that acts effectively as a smooth  infrared cutoff. A review of AdS/QCD models can be found in \cite{Brodsky:2014yha}. 

It is possible to use holographic models to calculate another hadronic property: the decay constant. 
The  non hadronic decay of mesons is represented 
as a transition from the initial  state to the hadronic vacuum.  For a meson  at radial excitation level $n$ with mass $m_n$ the decay constant is defined by the relation $ \langle 0 \vert \, J_\mu (0)  \,  \vert n \rangle = \epsilon_\mu f_n m_n \,, $
where $ J_\mu $ is the gauge current,  $ \epsilon_\mu $ the polarization and there is no implicit sum over $n$. Note that one finds other  definitions for the decay constants, involving different factors of the mass, in the literature. 

The two point function has a spectral decomposition  in terms of masses and   decay constants 
 of the states:  
 \begin{equation}
\Pi (p^2)  = \sum_{n=1}^\infty \, \frac{f_n^ 2}{(- p^ 2) - m_n^ 2 + i \epsilon} \,. 
\label{2point}
\end{equation} 
Calculating the left hand side of this equation using holography, one can find the mass and decay constant spectra \cite{Karch:2006pv,Grigoryan:2007my}.

In the finite temperature case, the particle content of a theory is described by the thermal spectral function, that is the imaginary part of the retarded Green's function. 
The quasi-particle states appear as peaks that decrease as the temperature or the density of the medium increase.  The limit of the spectral function when $T$ and $\mu$ vanish is a sum of delta peaks with coefficients proportional to the square of the decay constants: $ f_n^2 \, \delta ( - p^2 - m_n^2 ) $, arising from the imaginary part of eq.(\ref{2point}).  So, the decay constants control the amplitude of the delta function peaks that appear in the spectral function at zero temperature. When the temperature is raised, the peaks are smeared. The height and the width of the peaks become finite. But a consistent  extension to finite temperature must take into account the zero temperature behavior of the decay constants. 
That is why it is important to use  a model that provides nice fits for the decay constants when one wants to find a reliable picture of the thermal spectral function.

 \begin{table}[h]
\centering
\begin{tabular}[c]{|c||c||c||c|}
\hline 
\multicolumn{4}{|c|}{  Bottomonium  data   } \\
\hline
 &  Masses (MeV)   & $ \Gamma_{V \to e^+e^-} $ (keV) &   Decay constants (MeV) \\
\hline
$\,\,\,\, 1S \,\,\,\,$ & $ 9460.3\pm 0.26 $ &$ \,\,\,\, 1.340 \pm 0.018 \,\,\,\, $ & $ 715.0 \pm 2.4 $ \\ 
\hline
$\,\,\,\, 2S \,\,\,\,$ & $ 10023.26 \pm 0.32 $ & $ \,\,\,\,  0.612 \pm 0.011\,\,\,\,$  & $ 497.4 \pm 2.2 $  \\
\hline 
$\,\,\,\,3S \,\,\,\,$ & $ 10355.2 \pm 0.5 $ & $ \,\,\,\, 0.443 \pm 0.008 \,\,\,\, $ & $ 430.1  \pm 1.9 $ \\ 
\hline
$ \,\,\,\, 4S  \,\,\,\,$ & $ 10579.4 \pm 1.2 $ & $ \,\,\,\, 0.272 \pm 0.029 \,\,\,\,$  & $ 340.7  \pm 9.1 $ \\
\hline
\end{tabular}   
\caption{Experimental masses and electron-positron widths from \cite{Agashe:2014kda} and the corresponding decay constants for the Bottomonium S-wave resonances.  }
\end{table}

 In order to illustrate the behavior of the decay constants we show  on table {\bf I  } the experimental values of masses and electron positron decay widths $\Gamma_{ V \to e^+ e^ - } $ for botomonium vector meson  $ \Upsilon $, made of  a bottom quark anti-quark pair and  for the first three radially excited S-wave resonances.  We also show  the associated decay constants, with the corresponding uncertainties.
 The experimental values for masses and decay widths are taken from ref. \cite{Agashe:2014kda}. 
The decay constant of a vector meson state is related to it's mass and width by \cite{Hwang:1997ie}:
\begin{equation} 
f_{_V}^ 2 \,=\, \frac{3 m_{_V} \Gamma_{ V \to e^+ e^ - }  }{4 \pi \alpha^ 2 c_{_V}}\,,
 \label{decay-widths}
 \end{equation} 
 \noindent where $\alpha = 1/137 $ and $c_{_V} $ is  $c_{\Upsilon} = 1/9$.   
 The decay constants decrease monotonically with the radial excitation level. 
 A similar behavior is observed for charmonium vector states\cite{Braga:2015jca}.    
  In contrast, the soft wall model as originally formulated leads to decay constants for radial excitations of a vector meson that are degenerate
The hard wall model provides decay constants that increase with the excitation level. 
 The alternative version of the soft wall model developed in ref.  \cite{Braga:2015jca} is  consistent with the observed behavior.  The decay constants are obtained from two point correlators of gauge theory operators  calculated at a  finite value $ z = z_0$  of the radial coordinate of AdS space.   This way an extra energy parameter 
  $ 1/z_0 $, associated with  an ultraviolet (UV)  energy scale, is introduced  in the model. 
  
  In the  subsequent article of ref. \cite{Braga:2016wkm} this model was applied to the finite temperature and zero chemical potential case. A nice picture for the dissociation of $1S, 2S$ and $ 3S$ states of botomonium emerged, consistent with previous results \cite{Adare:2014hje}.   The purpose of the present letter  is to extend the study of bottomonium  vector meson dissociation for the case of finite density.     
  
Heavy mesons have also been discussed in the context of holography in some interesting articles as, for example, refs. \cite{Asakawa:2003re,Hong:2003jm,Kim:2007rt,Fujita:2009wc,Noronha:2009da,Fujita:2009ca,Branz:2010ub,Gutsche:2012ez,
Afonin:2013npa,Hashimoto:2014jua,Liu:2016iqo,Liu:2016urz}.   However the picture for  the dissociation of   $ 1S$ and $2S$   states of  bottomonium in a medium with finite temperature and density that we show here  is not yet present in the literature. 
  
 The article is organized as follows. In section II we  explain the relation between the decay constants and the spectral function peaks. We also review the model of refs. 
  \cite{Braga:2015jca,Braga:2016wkm} and  explain the reason for using such a model with UV cut off 
  in the finite temperature and density case. In section III we  present an extension to finite chemical potential.  Then is section IV  we develop the calculation of the vector meson spectral function using the membrane paradigm.  In section V we analyze the bottomonium thermal spectrum as a function of $T$ and $\mu$ and discuss the results obtained.

\section{Holographic model for decay constants}

\subsection{Decay constants in the soft wall}

In the soft wall model \cite{Karch:2006pv}  vector mesons are described by a vector field  $V_m = (V_\mu,V_z)\,$ ($\mu = 0,1,2,3$), 
assumed to be dual to the gauge theory current $ J^\mu = \bar{\psi}\gamma^\mu \psi \,$. The action is:
\begin{equation}
I \,=\, \int d^4x dz \, \sqrt{-g} \,\, e^{- \Phi (z)  } \, \left\{  - \frac{1}{4 g_5^2} F_{mn} F^{mn}
\,  \right\} \,\,, 
\label{vectorfieldaction}
\end{equation}
\noindent where $F_{mn} = \partial_mV_n - \partial_n V_m$ and $\Phi = k^2z^2   $ is the soft wall dilaton background, that plays the role of a smooth infrared cut off and  $k$ is a constant representing the mass scale.  

The  background geometry of the model is anti-de Sitter $AdS_5$ space,  with metric
\begin{equation}
 ds^2 \,\,= \,\, e^{2A(z)}(-dt^2 + d\vec{x}\cdot d\vec{x} + dz^2)\,,
\end{equation}
\noindent where $ A(z) = -log(z/R) $ and  $ (t,\vec{x})\in \mathcal{R}^{1,3},~z\in (0,\infty)$. 

One uses  the gauge  $V_z=0$. The boundary values of the other  components of the vector field:  
$ V^0_{\mu}(x) =
\lim_{z\to 0} V_\mu (x,z) , \mu = 0,1,2,3\, $ are assumed to be, as in  AdS/CFT correspondence,  sources of  correlation functions of the boundary current operator    
\begin{equation}
 \langle 0 \vert \, J_\mu (x) J_\nu (y) \,  \vert 0 \rangle \, =\, \frac{\delta}{\delta V^{0\mu}(x)} \frac{\delta}{\delta V^{0\nu}(y)}
 \exp  \left( - I_{on shell} \right)\,,
 \label{2pointfunction}
\end{equation}
\noindent where the on shell action is given by the boundary term: 
 \begin{equation}
I_{on \, shell }\,=\, - \frac{1}{2 {\tilde g}_5^2}  \, \int d^4x \,\,\left[  \frac{e^{- k^2 z^ 2  }}{z} V_\mu \partial_z V^ \mu 
\right]_{_{ \! z \to 0 }}
 \,.
\label{onshellaction}
\end{equation}
For convenience we introduced $ {\tilde g}_5^2 =  g_5^2 /R $, the relevant dimensionless coupling of the vector field.
One can write the on shell action in momentum space and decompose the field as 
 \begin{equation} 
 V_\mu (p,z) \,=\, v (p,z) V^0_\mu ( p ) \,,
 \label{Bulktoboundary}
\end{equation}  
 \noindent where    $ v (p,z)   $ is called bulk to boundary propagator and satisfies the equation of motion:
\begin{equation}
\partial_z \Big( \frac{ e^{-k^ 2 z^ 2 }} { z}  \partial_z v (p,z) \Big) + \frac{p^ 2 }{z} e^{-k^ 2 z^ 2 }  v (p,z) \,=\, 0   \,.
\label{BulktoboundaryEOM}
\end{equation}
The factor $ V^0_\mu ( p )$  works as the source of the correlators of gauge theory currents, so one imposes the boundary condition:   $ v (p, z=0) = 1\,$.
On the other hand,  the two point function in momentum space is related to the current-current correlator   by: 
 \begin{equation}
  \left( p^2 \eta_{\mu\nu} -  p_\mu p_\nu   \right) \, \Pi ( p^2 ) 
\, =\, \int d^4x \,\, e^{-ip\cdot x} \langle 0 \vert \, J_\mu (x) J_\nu (0) \,  \vert 0 \rangle  \, . 
\label{correlatorand2pointfunction}
\end{equation}  
 The  two point function is expressed as: 
 \begin{equation}
  \Pi ( p^2 )  \, =\,   \frac{1}{{\tilde g}_5^ 2 \, (-p^ 2) } \left[  \frac{ e^{ -k^2 z^2 }  \,  v (p,z) \partial_z v (p,z)  }{  \, z  } 
  \right]_{_{ \! z \to 0 }} \, ,
 \label{hol2point}
\end{equation}      
 and has the spectral decomposition shown in eq. (\ref{2point}), in terms of masses and   decay constants.  The result for the decay constants, following this soft wall approach is \cite{Karch:2006pv}:
 \begin{equation}
 f_n    \, =\,  
  k \sqrt{ 2}  / {\tilde g}_5  \,.
 \label{decayconstants}
\end{equation}  
\noindent This means that in the soft wall model all the radial excitations of a  vector meson have the same decay constant. This result contrasts with the decrease with excitation level $n$  obtained from experimental data shown on table {\bf I}.

\subsection{Alternative description of the decay process}

Now let us review the model proposed in ref. \cite{Braga:2015jca}, that leads to decay constants decreasing with excitation level.
The original soft wall model contains only one dimensionfull parameter:  $k$,  introduced in the dilaton background.    The proposal of ref. \cite{Braga:2015jca}  is   to incorporate in the model the effect of the interactions that govern the transition processes 
from a one hadron state to the hadronic vacuum, corresponding to the (non hadronic) decay of a vector meson.  One additional dimensionfull parameter in introduced by calculating holographically the operator product of currents of eq. (\ref{2pointfunction})  at a finite  location $ z = z_0$ of the radial coordinate.  The new parameter, namely $ 1/z_0 $,  corresponds to an (ultraviolet) energy  scale. 
It is worth to mention that a similar approach of introducing an ultraviolet cutoff in the soft wall model was considered before in ref. \cite{Evans:2006ea} for light vector mesons.
 
 The bulk to boundary propagator is written as a solution of the equation of motion (\ref{BulktoboundaryEOM})  but excluding the region $0 < z < z_0$. The solution is divided by a constant (in the $z$ coordinate) so as to  satisfy by construction the new boundary condition:
$$v(p,z_0) = 1. $$
Namely: 
 \begin{equation} 
 v (p,z ) \, = \, \frac{ U (p^ 2/ 4k^ 2 , 0, k^2 z^ 2 ) }{U (p^ 2/ 4k^ 2 , 0, k^2 z_0^ 2 )}\,,
 \label{bulktoboundary2}
\end{equation}  
\noindent where $U(a,b,c)$ is the Tricomi function. Using   the new on shell action given by:
 \begin{equation}
I_{on \, shell }\,=\, - \frac{1}{2 {\tilde g}_5^2}  \, \int d^4x \,\,\, \frac{e^{- k^2 z^ 2  }}{z} V_\mu \partial_z V^ \mu 
{\Big \vert}_{_{ \! z \to z_0 }} 
 \,,
\label{onshellaction2}
\end{equation}

\noindent one calculates the two point function from the on shell action at $z = z_0$:
\begin{equation}
  \Pi ( p^2 )  \, =\,   \frac{1}{{\tilde g}_5^ 2 \, (-p^ 2) }  \frac{ e^{ -k^2 z^2 }  \,  v (p,z) \partial_z v (p,z)  }{  \, z  } 
  {\Big \vert}_{_{ \! z \to z_0 }} 
   \, .
 \label{hol2pointnew}
\end{equation} 

\noindent     Then, using the new bulk to boundary propagator of eq. (\ref{bulktoboundary2}) and a  recursion relation for the Tricomi functions one  gets:
 \begin{equation} 
 \Pi ( p^ 2)  \, = \, \frac{1}{2 {\tilde g}_5^ 2 }\frac{ e^ {-k^ 2 z_0^ 2} U (1 + p^ 2/ 4k^ 2 , 1, k^2 z_0^ 2 ) }{U (p^ 2/ 4k^ 2 , 0, k^2 z_0^ 2 )}\,.
 \label{correlator2}
\end{equation}

The masses and decay constants are obtained  from the analysis of the poles of the two point function in the momentum variable $p^2$.  The singularities come from the zeroes of the denominator of eq. (\ref{bulktoboundary2}).  In a neighborhood of $ p^ 2 = p^ 2_n$ (simple pole) one can approximate  the two point function by 
\begin{equation}
\lim_{p^ 2 \to p^ 2_n} \Pi (p^2)   \approx \frac{f_n^ 2}{(-p^ 2)  + p_n^ 2 } \,.
\label{2pointlimit}
\end{equation} 
\noindent One associates the coefficients of the approximate expansion near the pole with the decay constant $f_n$ in analogy with the exact expansion shown in eq. (\ref{2point}). Thus one finds  the masses from the localization of the poles of the two point function and the decay constants from the
corresponding  coefficient. The parameters used for bottomnium in ref.\cite{Braga:2015jca}  are:  
$   k = 3.4 $   GeV ; \,  $ 1/z_0 = 12.5 $  GeV. 
The energy scale $1/z_0$ is flavor independent and represents a characteristic energy of the decay process. In the non hadronic decay a very heavy vector meson annihilates into light leptons. So, the ultraviolet scale is of order of the heavy  bottomonium masses.    The parameter $k$ represents the heavy quark mass.    The decay constants decrease with excitation level, as observed from experiments.

\section{ Finite density plasma } 
   
The holographic model proposed in  ref. \cite{Braga:2015jca} was extended  to finite temperature in \cite{Braga:2016wkm}. 
Now let us extend it to heavy vector mesons in a plasma at finite density. The action has  again the form of eq. 
(\ref{vectorfieldaction})   but the metric is now that of a charged black hole as for example in refs. \cite{Colangelo:2010pe,Colangelo:2011sr,Colangelo:2012jy} 
\begin{equation}
 ds^2 \,\,= \,\, \frac{R^2}{z^2}  \,  \Big(  -  f(z) dt^2 + \frac{dz^2}{f(z) }  + d\vec{x}\cdot d\vec{x}  \Big)   \,,
 \label{MinkoviskyMetric}
\end{equation}
with
\begin{equation}
f (z) = 1 - \frac{z^ 4}{z_h^4}-q^2z_{h}^{2}z^4+q^2z^6 
\end{equation}
The parameter $q$ is proportional to the charge of the black hole and  $z_h$ is the position of the horizon, defined by the condition $f(z_h)=0$. The Hawking temperature of the black hole is determined by the condition that there is no conical singularity at the horizon in the Euclidean version of metric (\ref{MinkoviskyMetric}).  This condition implies that the temporal coordinate $t$ must have a period of $\Delta t = 4 \pi / \vert f'(z)\vert_{(z=z_h)} $.  In the present case the gauge theory lives in the slice $z = z_0$ , so one must consider a local time variable $\tau = t \sqrt{f(z_0)} $ that leads to a transverse metric that is of Minkowski type, up to a conformal  factor. The temperature of the gauge theory is the inverse of the period of the local time $\tau$. So
\begin{equation} 
T = \frac{  \frac{\vert  f'(z)\vert_{(z=z_h)}}{4 \pi  }}{ \sqrt{ f (z_0) }}  = \frac{ \frac{1}{\pi z_h}-\frac{  q^2z_h ^5}{2 \pi }}{  \sqrt{ 1 - \frac{z_0^4 }{z_h^ 4 }- q^{2}z_{h}^{2}z_{0}^{4}+q^2z_{0}^{6}  \,}}\,.
\label{temp}
\end{equation}

The quark chemical  potential $\mu$ is related to $q$ as follows. In the gauge theory side of the duality $\mu$ appears as a constant parameter that enters the lagrangian multiplying the quark density 
$  \bar{\psi}\gamma^0 \psi \,$.  
So it corresponds  to the source of the correlators of this operator, that is the  field $V_0$.
In order to identify the chemical potential, one considers a particular solution for the vector field    $ V_m $ that has only one non vanishing component: $  V_0 = A_0 (z) $  ($V_z =0, V_i = 0$). Assuming as in 
\cite{Colangelo:2010pe,Colangelo:2012jy} that the relation between $q $ and $\mu$ is the same as in the case of is no soft wall background ($ k = 0$ ), the solution for the time component of the vector field is:
 $ A_{0}(z)=c - qz^2 $, where $c$ is a a constant. In the present model, the sources are the fields at $z=z_0$, so one imposes $A_{0}(z_0) = \mu $. Together with the condition $A_{0}(z_h)=0$ this implies
\begin{equation}
\mu = q z_h^2 - q z_0^2
\end{equation} 
The values of the dimensionless combination $ Q = qz_{h}^{3} $ must be in the interval
$0\leq Q \leq \sqrt{2}$.

An important point to be remarked is that in  AdS/CFT the sources of correlation functions correspond to non-normalizable solutions of the supergravity fields. This happens because these solutions have a non vanishing limit when $z \to 0$.
 The metric is singular at $z=0$ and the normalization involves an integral of the square of the function times a factor $1/z$. So the integral is singular at $z=0$.  
  In the model discussed here  $z_0 < z <  z_h$.
 The fields have non vanishing values at the boundary $z = z_0$,  so they work as sources. However there is no singularity in the normalization integral.  
 So, in contrast to AdS/CFT ,  the sources are not non-normalizable fields.  
  This happens with both the solution $A_0(z) $ that acts as the source of the quark density
  and  the solution (\ref{bulktoboundary2}) that is the source of the current correlators in momentum space.

\section{Spectral Function}
  
The spectral function of  bottomonium  can be calculated using the membrane paradigm \cite{Iqbal:2008by} (see also \cite{Finazzo:2015tta}). Let us review how to extend this formalism to a vector field $V_\mu$ in the bulk, dual to the electric  current operator $J_{\mu}$. We consider a general black brane background  of the form
\begin{equation}\label{metric3}
ds^{2}=-g_{tt}dt^2+g_{zz}dz^{2}+g_{x_1 x_1}dx^{2}_{1}+g_{x_2 x_2}dx^{2}_{2}+g_{x_3 x_3}dx^{2}_{3}\,,
\end{equation}
where we assume the boundary is at a position  $z=z_0$ and  the position  of the horizon of the black hole, $z_{h}$, is given by the root of $g_{tt}(z_{h})=0$. The bulk action for the gauge fields  is
\begin{equation}\label{Sigma2}
S=-\int d^{5}x \, \sqrt{-g}\frac{1}{4 g_5^2 \, h(z)}F^{m n}F_{m n}\,,
\end{equation}
where $h(z)$ is  a z-dependent coupling. The equation of motion  that follows from the bulk action is:
\begin{equation}\label{eqms}
\partial^{m} \left( \frac{\sqrt{-g}}{h(z)}  F_{m n } \right)=0 \,.
\end{equation}
The conjugate momentum to the gauge field with respect to a foliation by constant z-slices is given by:
\begin{equation}\label{momentum}
j^{\mu}=-\frac{1}{h(z)}\sqrt{-g}F^{z \mu} \,.
\end{equation}
One  can consider solutions for the vector field that do not depend on the coordinates  $x_1$ and $x_2$. Then eq. (\ref{eqms})  can be separated in  two different channels: a longitudinal one  involving fluctuations along $(t, x_3)$ and a transverse channel involving fluctuations along  spatial directions  $(x_1,x_2)$. 
Using eq. (\ref{momentum}),  the components $t$, $x_3$ and $z$ of  eq. (\ref{eqms}) can be written respectively as
\begin{equation}\label{Maxwell1}
-\partial_{z}j^{t}-\frac{\sqrt{-g}}{h}g^{tt}g^{x_3 x_3}\partial_{x_3}F_{x_3t}=0\,,
\end{equation}
\begin{equation}\label{Maxwell2}
-\partial_{z}j^{x_3}+\frac{\sqrt{-g}}{h}g^{tt}g^{x_3 x_3}\partial_{t}F_{x_3t}=0\,,
\end{equation}
\begin{equation}\label{Maxwellz}
\partial_{x_3}j^{x_3}+\partial_t j^{t}=0\,.
\end{equation}
 From the Bianchi identity one finds the relation:
\begin{equation}\label{Bianchi}
\partial_{z}F_{x_3t}-\frac{h(z)}{\sqrt{-g}}g_{zz}g_{x_3 x_3}\partial_{t}j^{z}-\frac{h(z)}{\sqrt{-g}}g_{tt}g_{x_3 x_3}\partial_{x_3}j^{t}=0\,.
\end{equation} 
Now, one can define a z-dependent "conductivity" for the longitudinal channel given by:
\begin{equation}\label{longitudinal}
\bar{\sigma}_{L}(\omega,\vec{p},z)=\frac{j^{x_3}(\omega,\vec{p},z)}{F_{x_3 t}(\omega,\vec{p},z)}\,.
\end{equation} 
Taking a derivative of the equation above with respect to $z$, using  eqs.  (\ref{Maxwell1}),  (\ref{Maxwellz}) and (\ref{Bianchi}) and considering a plane wave solution with momentum  $p=(\omega,0,0,p_3)$ one finds
\begin{equation}\label{Membrane}
\partial_{z}\bar{\sigma}_{L}=-i\omega\sqrt{\frac{g_{zz}}{g_{tt}}}\left[ \Sigma(z)-\frac{\bar{\sigma}_{L}^{2}}{\Sigma(z)} \left( 1-\frac{p_3^2}{\omega^2} \frac{g^{x_3 x_3}}{g^{tt}}          \right)    \right] \,,
\end{equation}
where 
\begin{equation}\label{Sigma}
\Sigma (z)=\frac{1}{h(z)}\sqrt{\frac{-g}{g_{zz}g_{tt}}}g^{x_3 x_3} \,.
\end{equation}
Following a similar procedure for the transverse channel one finds  \cite{Iqbal:2008by}
 the equation for $\bar{\sigma}_{T}$:
\begin{equation}\label{Membrane2}
\partial_{z}\bar{\sigma}_{T}=i\omega\sqrt{\frac{g_{zz}}{g_{tt}}}\left[ \frac{\bar{\sigma}_{L}^{2}}{\Sigma(z)} - \Sigma(z)\left( 1-\frac{p_3^2}{\omega^2} \frac{g^{x_3 x_3}}{g^{tt}}          \right)    \right] \,.
\end{equation}
Note that in zero  momentum limit, both flow equations have the same form
 \begin{equation}\label{MembraneF}
\partial_{z}\bar{\sigma}=i\omega\sqrt{\frac{g_{zz}}{g_{tt}}}\left[ \frac{\bar{\sigma}^{2}}{\Sigma(z)} - \Sigma(z)    \right] \,,
\end{equation}    
for $\bar{\sigma}= \bar{\sigma}_{T}=\bar{\sigma}_{L}$.
The Kubo formula $ \sigma(\omega)= i G_{R} ( \omega) / \omega\,,$    
where $\sigma$ is  the AC  conductivity, motivates one define: 
\begin{equation}\label{AC}
\sigma(\omega) =-\frac{G_{R}(\omega)}{i \omega}\equiv \bar{\sigma}(\omega,z_0)\,.
\end{equation} 
Notice that $Re$ $\sigma(\omega)=\rho(\omega)/\omega$, where $ \rho(\omega) \equiv Im \ G_{R}(\omega)$ is the spectral function.  

In order to apply the membrane paradigm to the model of the previous section, we use the metric (\ref{MinkoviskyMetric}) and $ h(z) = e^{k^2z^2 } $
 in the flow equation  (\ref{MembraneF}). One finds
\begin{equation}\label{MembraneF2}
\partial_{z}\bar{\sigma}(\omega,z)= \frac{i\omega}{f(z)\bar{\Sigma}(z)} \left[ \bar{\sigma}(\omega,z)^{2}-\bar{\Sigma}(z)^{2}  \right] \,,
\end{equation}      
with $\bar{\Sigma}(z)=e^{-k^2z^{2}}/z$. Requiring regularity at the horizon, one obtains the following condition: $ \bar{\sigma}(\omega,z_h)=\bar{\Sigma}(z_{h})\,.$ The spectral function is obtained from the relation   
\begin{equation}\label{spectralfunction}
\rho(\omega)\equiv Im \ G_{R}(\omega)= \omega  Re \  \bar{\sigma}(\omega,z_0)\,.
\end{equation}

\begin{figure}[h]
\label{g6}
\begin{center}
\includegraphics[scale=0.3]{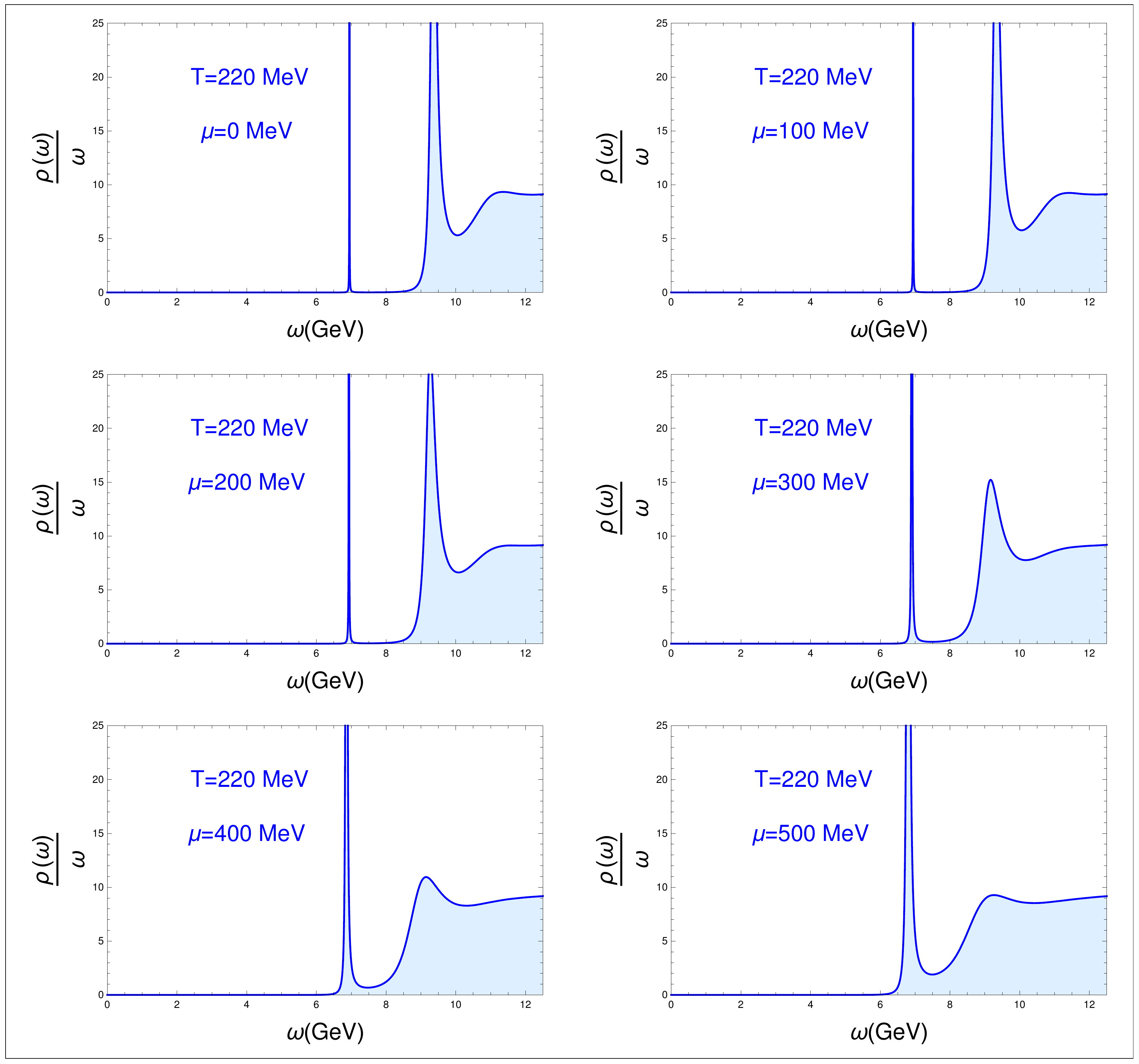}
\end{center}
\caption{Spectral functions for $T=$ 220 MeV at 6 representative  values of $\mu$ }
\end{figure}
 \begin{figure}[h]
\label{g6}
\begin{center}
\includegraphics[scale=0.3]{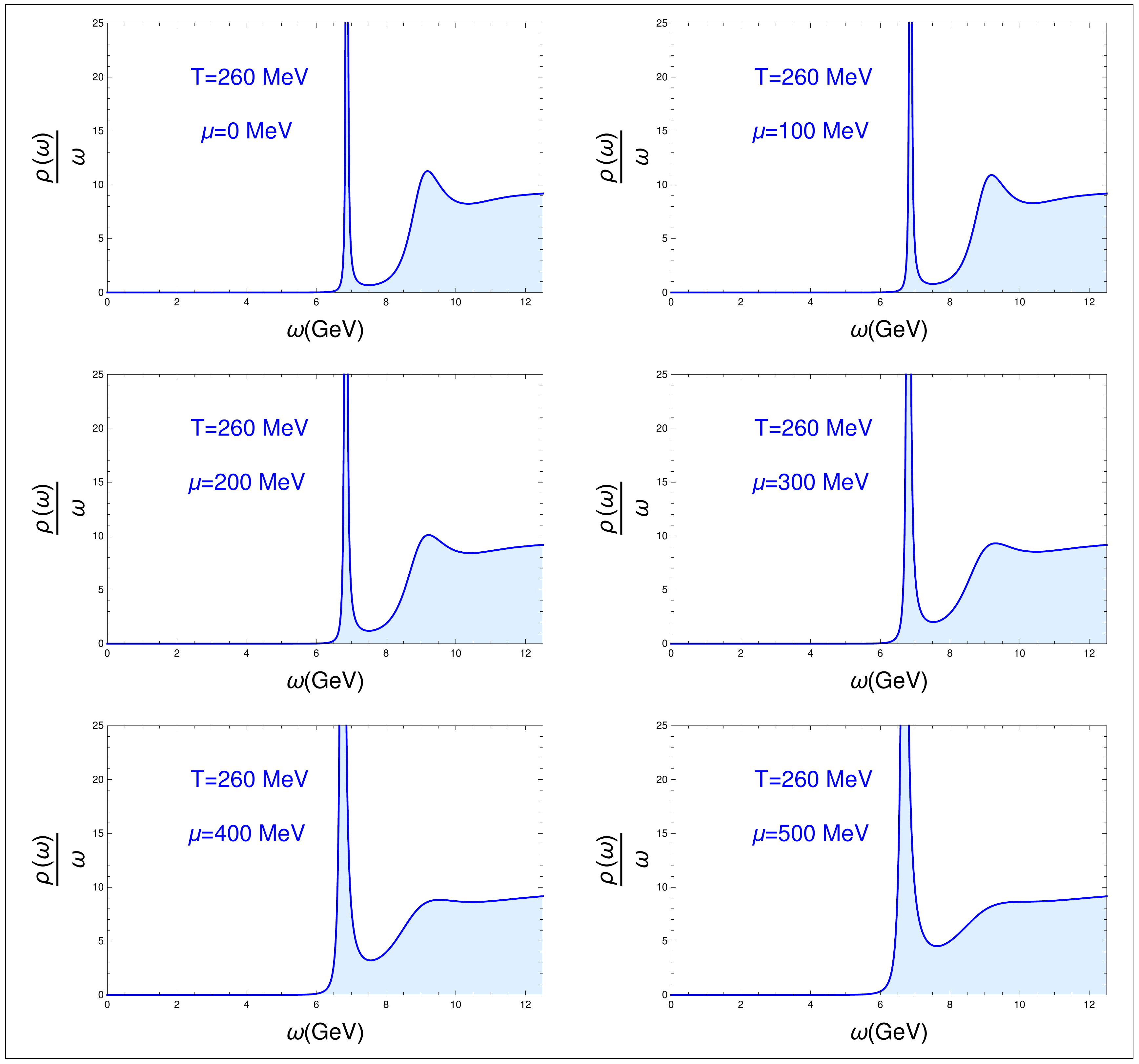}
\end{center}
\caption{Spectral functions for $T=$ 260 MeV showing the dissociation of the $1S $ state at larger values of $\mu$ }
\end{figure}

\section{Results and Discussion}        
    
  The procedure to obtain  the spectral function for bottomonium  is to solve  numerically equation (\ref{MembraneF2})  with the boundary conditions described in the previous section. For the model parameters, we use the values  obtained in ref. \cite{Braga:2015jca} from the fit of the experimental data at zero temperature and chemical potential: $ 1/z_0 = 12.5  \,{\rm GeV}  ;  \,  k = 3.4 \,  {\rm GeV  } $.
 The spectral functions (\ref{spectralfunction}),  calculated for different temperatures and chemical potentials grow, for large frequencies $\omega $,  as: $ \rho \sim \omega $. So, we analyzed the behavior of re-scaled spectral function: $ \bar{\rho} (\omega)  = \frac{ \rho(\omega ) }{\omega } \,,$ in the intervall  $ 0 < \omega < 12.5 $ GeV (UV cut off). 
    
  \begin{figure}[h]
\label{g6}
\begin{center}
\includegraphics[scale=0.3]{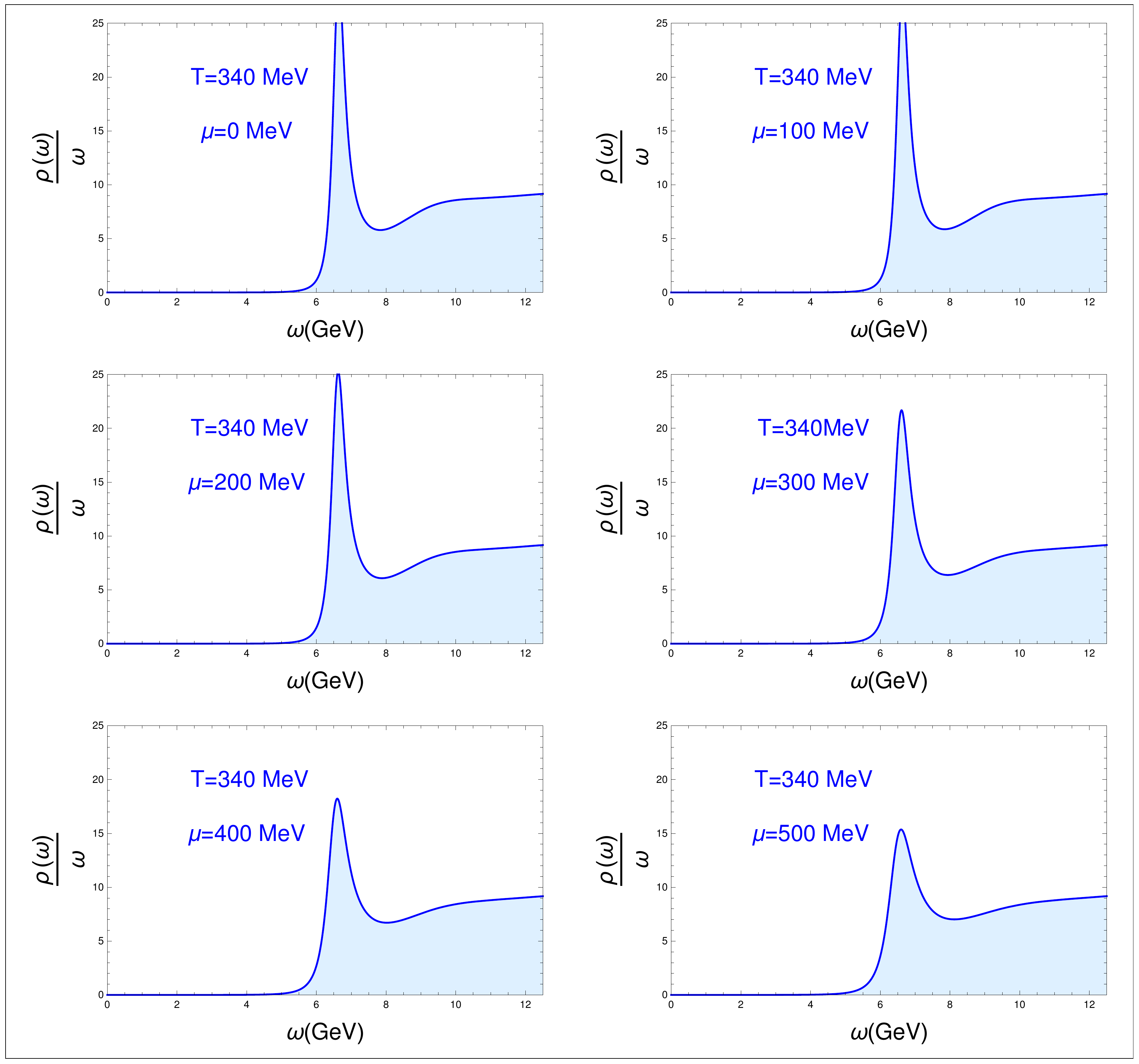}
\end{center}
\caption{Spectral functions for $T=$ 340 MeV at 6 representative  values of $\mu$ }
\end{figure}

 The critical confinement/deconfinement temperature of this model -- with ultraviolet cut off -- was estimated for the case of zero chemical potential in  ref. \cite{Braga:2016wkm} as $T_c = 191$ MeV (see also \cite{Witten:1998zw,Hawking:1982dh,Herzog:2006ra,BallonBayona:2007vp} ).
 We  start the analysis with a temperature slightly above, showing   in figure {\bf 1 } the case $T= 220 $ MeV for six  different values of the chemical potential from $ \mu = 0 $ to $ \mu = 500 $ MeV.    One can see on the first  panel the peaks corresponding to the $1 S$ and $2S $ states and a very small peak corresponding to the $3S$ state. Then, raising the value of $\mu $ the second peak dissociates while the first one develops a larger width.

Then in figure {\bf 2 } we show the case of $ T= 260 $ MeV where one sees a clear reduction of the second quasi-particle peak associated with the $2 S$ state at $\mu =0$. Increasing $\mu$ one sees that this state is completely dissociated in the medium at $\mu = 500 $ MeV. 

In figure {\bf 3 } we present the $T = 340 $ MeV case.  One can see that at $\mu = 0$ only the first peak corresponding to the $1 S $ state is present. Then increasing $\mu $ the quasi-particle peak has a noticeable decrease, corresponding to dissociation in the medium. 

It is interesting to note that the results shown appreciable variation in the particle content of the plasma in the quark chemical potential interval  $  0 <  \mu  <  500 $ MeV.
This finding is consistent with the quantitative expectations for the QCD phase diagram \cite{Fukushima:2010bq}. 
      
A remarkable fact that can be seen form the plots is that the effect of the finite density is much  stronger for  the higher excited states. For example, at $ T = 260 $ MeV the $2S $ is present at $ \mu = 0$ but it disappears at $ \mu = 500 $ MeV. 
For the third excited state there is just some small trace at $ T = 220 $ MeV and $ \mu = 0$. 

Let us finally mention that this AdS/QCD model with UV cut off can be used for other hadronic states. It was applied for charmonium states in ref. \cite{Braga:2015jca}  using the same UV cut off 
$ 1/z_0 = 12.5  \,{\rm GeV} $ but with the soft wall parameter $  k_c = 1.2 \,  {\rm GeV  } $. It can be applied for light mesons also. In particular, for the $\rho$ meson one uses the same UV cut off but $ k_\rho = 0.39 $ GeV, finding masses of 776 MeV, 1098 MeV and 1345 MeV for the first radial states. 
Comparing these different hadronic cases one notices that $ k$ is related to both the quark mass and the string tension while $ 1/z_0 $ is a fixed UV scale. 

The motivation for introducing the UV cut off in AdS space  is twofold: to take the quark masses into account and to describe the decrease in decay constants with radial excitation number. In particular, this second property is not reproduced in other models with different backgrounds. For example, in the D4-D8 brane model one finds decay constants increasing with excitation level\cite{BallonBayona:2009ar}.

\noindent {\bf Acknowledgments:}  We thank  Eduardo Fraga for important discussions.   N.B. is partially supported by CNPq and L. F.  is supported by CAPES.

 \end{document}